\documentstyle[preprint,aps]{revtex}
\def\lsim{\mathrel{\rlap{\lower4pt\hbox{\hskip1pt$\sim$}}
    \raise1pt\hbox{$<$}}}         %less than or approx. symbol
\def\gsim{\mathrel{\rlap{\lower4pt\hbox{\hskip1pt$\sim$}}
    \raise1pt\hbox{$>$}}}         %greater than or approx. symbol
\begin{document}
\bibliographystyle{unsrt}
%\draft
\title{Nuclear Physics}
\author{ E.M. Henley}
\address{ Department of Physics and Institute for Nuclear Theory, 
University of Washington,\\
  Box 351560, Seattle, WA 98195}
\author{ and}
\author{ J.P. Schiffer}
\address{ Argonne National Laboratory, Argonne, IL 60439 and University of
 Chicago, Chicago, IL 60637\\}
\maketitle
\begin{abstract}
Nuclear Physics is the branch of physics that deals with the properties 
and structure of matter on the hadronic level.  In this article we 
review briefly the history of this field, which has a major role in the 
development of our understanding of nature.  We then proceed to give an 
outline of a current perspective of the field and of some of the issues 
that are now on its frontiers.

\end{abstract}

\section{Introduction}

\subsection{Historical Perspective}

 	As we are approaching the turn of the century we wish to review very
briefly the status of nuclear physics as it has evolved in the course of our
explorations of nature during this period. 

	Nuclear physics was born over 100 years ago with the discovery of
radioactivity by Becquerel and followed by the work of the Curies in
identifying the sources of the new radiations.  The discovery of electrons by
J.J. Thomson occurred approximately at the same time and the identification of
one of the radiations in radioactivity (beta rays) as electrons, pointed to a
close connection in this new realm of nature.  The nucleus as a small heavy
center of the atom, was only deduced in 1911 by Rutherford from the large
angles into which energetic alpha particles from a radioactive source were
scattered when incident on a foil.  The simplest nucleus was evidently that of
the hydrogen atom, the proton and hence thought to be ``elementary" along with
the electron.  The existence of a nucleus was a crucial feature of the Bohr
atom -- and thus central to the development of quantum mechanics.  Until the
discovery of the neutron in 1932 it was believed that the nucleus was made of
electrons and protons.  The nucleus of protons and neutrons, as we know it,
can be said to date from that time. 

	The first studies of nuclear reactions were carried out with alpha
particles from radioactive decay -- and it was quickly realized that to pursue
such studies required more intense sources of charged particles.  Several
classes of accelerators, to provide such energetic particles, were invented
and developed for the study of the nucleus in the 1930's, from the
electrostatic generators of high voltage by Cockroft and Walton and Van de
Graaff, to the cyclotron of Lawrence, to betatrons and their derivatives.
Techniques for detecting particles also started in this period from the early
scintillating screens of Rutherford to the counters and cloud chambers of the
1920's and 30's.  The early work on nuclear reactions quickly established the
size of the nucleus and thus the range of nuclear forces.  The electron
spectrum of beta decay led Pauli to the supposition that there had to be an
additional particle involved, the neutrino, and Fermi subsequently formulated
the correct theory of the beta decay process.  These early developments of
nuclear physics provided a rich case study for the application of the new
quantum theory. 

	One of the surprising features of the nucleus was the fact that
nuclear forces seemed to have a very short range.  In 1935 Yukawa postulated
exchange forces and a ``meson" as the carrier of the nuclear force that keeps
the neutrons and protons bound inside the nucleus.  Such a particle, of mass
between the electron and proton that was thought to be the meson, was found in
cosmic rays in 1937, but it was realized that its nuclear interactions were
too weak.  We now know that this was the muon and the pion was not discovered
until 1946.  The 1930's also saw the first realization by Bethe of how nuclear
reactions fuel the sun in converting hydrogen into the light elements. 

	Progress in nuclear physics was rapid after this start and the last 50
years have seen explosive growth in our understanding of the nucleus, its
constituents and its forces.  The internal symmetry between n-p 
(neutron-proton), n-n and p-p forces, now known as charge independence,
was already postulated in the 1930's and its implications explored by
Heisenberg, by Wigner and others.  This led to the concept of isospin
that has had a crucial role, both in particle and nuclear physics. 

	The realization in the late 1940's by M. Goeppert-Mayer and J.H.D.
Jensen that the structure of the nucleus can be understood in terms of a shell
model, as in atomic physics, came as a great surprise because of the high
density of the nucleus and the strength of nuclear forces, but the validity of
this description has been verified by many phenomena.  Its theoretical basis
came later.  Although heavy nuclei could be built up from nucleons, it was
realized that there must be simpler degrees freedom and descriptions.  There
were many such models, including a model of an almost degenerate Fermi gas of
nucleons, and a liquid drop model.  The dynamics of fission, discovered in
1939 by Hahn and Meitner, could be understood on the basis of such models. 
The large quadrupole deformation of some classes of nuclei led to a further
understanding of new degrees of freedom, and the dynamical collective model of
Bohr and Mottelson came out of this realization.  These degrees of freedom,
primarily collective rotations of non-spherical nuclei as well as vibrations,
lead to a simplified description, particularly, of deformed nuclei.  When
combined with the shell model in a deformed potential, this work led to a
unified model of the nucleus.\cite{Boh}  An important degree of freedom
in nuclei was found to be that of ``pairing", the correlations between pairs of
nucleons coupled to zero spin, and the theoretical understanding has a close 
analogy with the BCS theory of superconductivity involving the pairing of
electrons.  More recently, a very successful model has emerged, algebraic
models of ``dynamic symmetries" named the interacting boson model.  Here the
degrees of freedom are those of a boson, primarily of zero spin.\cite{Ari} 
The unified model and the mean-field theories also permitted descriptions of
excited states and transition rates.  They also led to an understanding of
giant resonances seen in the excitation of a nucleus, be they monopole
(breathing mode), dipole, or higher multipoles. 

	In parallel with the vast improvement in our understanding of the
structure of nuclei was work on nuclear reactions.  Early work showed
pronounced resonances in nuclear reactions, particularly the absorption
of slow neutrons, indicating long-lived intermediate states.  This implied
many degrees of freedom and apparently a complicated process.  Considerable
simplification resulted from descriptions in which the description of the 
compound system through which a reaction proceeds is described in terms of 
the {\it average} properties of the resonances, without the detailed
consideration of their individual properties.  Reactions can be considered in
two limits, one proceeding through long-lived intermediate resonances whose
decay is independent of their mode of formation -- the simple expression for a
resonance was given by Breit and Wigner in the 1930's, followed by the
detailed treatment in reaction formalisms, particularly of Wigner in the
1940's and 50's.  The other limit developed in the 1960's is that of direct,
one step reactions, proceeding in a time comparable to the passage of the
projectile through the nuclear volume.  In this limit, the reaction may be
described by an effective ``optical" potential with a real and imaginary part.
Here the interaction can be described as a perturbation in a quasi-elastic
scattering process with the incident wave modified by the interaction but
continuing coherently in the outgoing channel.  Formalisms for the
descriptions of more complicated reactions between the two extremes were
filled in more slowly. 

	The 1950's and 60's also saw major new developments in beta decay, the
realm of the ``weak" interactions.  The ideas of Lee and Yang that parity need
not be conserved in beta decay was quickly followed by the work of C.S. Wu
showing that, indeed, parity conservation was not a valid symmetry in these
processes.  The study of nuclear beta decay has laid the foundations of the
Standard Model. 

	The 1970's and 80's saw the consolidation of the description of the 
nucleus with improved experimental techniques supporting the theoretical 
framework.  The use of computers aided this endeavor greatly.  The use of 
heavy ion beams (accelerated heavy nuclei) was expanded and new features of
nuclei were investigated.  The use of electron beams for mapping of nuclear
charge distributions, pioneered by Hofstadter, was extended in precision.  
Distributions of charge and magnetization densities and transition
probabilities were mapped out and this field was established as a powerful 
quantitative source of information about nuclei.  More generally, detection
techniques evolved considerably in resolution and in the capacity to handle
complex information.  

	New accelerators using superconductivity were developed, both linear
accelerators with superconducting rf cavities and cyclotrons with
superconducting magnets.  Also, the realization of the pion's role in the
nucleus lead to the construction of ``meson factories", where intense beams of
pions were produced for studies of nuclear properties.  Strange mesons
produced at high-energy accelerators were used to produce ``hypernuclei", a new
class of nuclei in which a long-lived baryon with a ``strange" quark is bound
along with neutrons and protons.

\subsection{ Present Perspective}

        Coming into the current decade, nuclear physics continues to address
the state of hadronic matter, which increasingly includes the structure of
hadrons as well as the larger many-body aspects of nuclei.  The field of 
nuclear physics at the end of the century encompasses a number of areas and in 
this article we will attempt to discuss briefly a few of the current thrusts 
and cite a few review articles that provide more details. 

	The hadrons are the simplest entities of strongly interacting matter
that can exist by themselves.  Their properties are well established 
experimentally, but the way they are constituted out of quarks and gluons is
still not well understood.  Recent experimental results have shown that the
spin of the nucleons, for instance, is not as simple as it seemed a few years
ago, but has contributions from the polarization of the QCD ``sea", from gluons
and from the possible angular momentum of quarks and gluons.  How protons and
neutrons -- the most stable hadrons -- interact with each other to form simple
nuclei has seen substantial progress.  Evidence is now quite conclusive that
simple two-body forces are insufficient to explain the properties of the
simple nuclei and that many-body forces are important. 

	The understanding of the structure of nuclei in terms of the shell
model and the various collective rotations, vibrations and excitations that
nuclei undergo has advanced in several directions.  In particular, new
detection techniques have helped unravel bands of states that correspond to
shapes that are deformed from spherical symmetry much more drastically than
previously observed -- suggesting a region of stability with 2:1 (major to 
minor) axis ratios.  These states appear to be rather pure and hardly mix at
all with the more spherical normal states.  Other advances in experimental
capabilities have allowed the approach of the limits of nuclear stability, the 
so-called drip line. 

        One of the aspects of QCD  that is not satisfactorily understood is
the concept of {\it confinement}, the fact that the constituents of hadrons
can never appear isolated by themselves.  At very high densities this
confinement is expected to break down to the extent that quarks can travel
freely within the region of high energy density.  This is presumably a state
that the universe passed through shortly after the Big Bang.  There will soon
be a new tool for investigating the state of matter at that time:  a large
collider of heavy (e.g., Au) ions is being constructed in which similar energy
densities should be reached.  A key symmetry that is broken in normal QCD,
that of {\it chiral symmetry}, may well be restored in this regime of energy
density.  Both the experimental undertakings at this facility and the
theoretical interpretations are a major challenge for the field in the coming
decade. 

        The crucial role of nuclear physics in fueling the stars has been 
recognized since the early work of Bethe, who showed how stars are powered by
fusion reactions, and later of Fowler and coworkers who developed the 
understanding of the processes responsible for the formation of elements. 
Specific nuclear properties play key roles in the Big Bang, in the energy
production in our Sun and in other stars, and in all the nucleosynthetic
processes of the Universe.  This intimate relationship is beautifully
illustrated by the fact that the properties of the lightest neutrinos have
been enormously clarified by the theoretical interpretation of experiments
that searched for the nuclear reactions that these neutrinos induce on earth. 

        Finally, not only has the field depended critically on developing a
large variety of experimental and theoretical techniques, but these techniques
have in turn served society in a number of ways -- the prevalence of nuclear
medicine being a prominent example.

\section{Hadron Physics}

	The smallest entities of accessible strongly-interacting matter in the
world are hadrons, either baryons that are aggregates of three quarks or
mesons that are made from quark-antiquark pairs.  The most stable baryons, the
protons and neutrons, are the major constituents of atomic nuclei, and the
lightest meson is the pion.  Understanding the structure of hadrons and how
the properties of these particles arise from QCD is a major interest of
nuclear physics.  This interest follows two paths.  One concerns the
properties of families of hadrons as they exist freely, to accurately
characterize the members of the rich hadron spectrum in mass and decay
properties and reflect the structure that arises from QCD.  The other is to
understand how these properties change when the hadrons are immersed in 
nuclei or nuclear matter.

\subsection{Pions}

	Among the mesons, the lightest and most important one is certainly the
pion.  Thus, it is no accident that its properties, production, and
interactions with nucleons and nuclei have received considerable attention in
the past and again at the present time. 

    	Because quantum chromodynamics (QCD) is the underlying theory of
hadronic interactions, there have been many models built on one aspect
or another of the theory.  A particularly important symmetry of QCD,
which is almost preserved at low energies, is chiral invariance or the
symmetry between left- and right-handedness.  This symmetry is
incorporated into the most recent treatments of few-body problems, with
the use of low-energy effective theories, such as chiral perturbation
theory, first introduced by Weinberg many years ago.\cite{Wein}  Here
an effective Lagrangian is constructed which incorporates all terms
allowed by the symmetries of QCD.  In QCD with massless up (u), down (d)
and strange (s) quarks, the theory satisfies chiral invariance.  This
leads to both conserved vector and axial vector currents and to parity
doublets.  Since the axial current is not conserved in nature and parity
doublets are not observed, one assumes that spontaneous symmetry
breaking leads to the eight Goldstone (almost massless) pseudoscalar
bosons.  The finite quark masses also break the symmetry somewhat, and
this leads to the non-vanishing pion and other light pseudoscalar meson
masses. 

    	The approximate chiral invariance is incorporated in all low
energy effective theories.  Chiral perturbation theory is a low-energy
theory with a systematic expansion around the chiral limit in powers of
$m_{\pi} / \Lambda$ and $momenta/\Lambda$, where $\Lambda$ is a
quantum chromodynamics (QCD) scale, of the order of 1 GeV.  Because it is
an effective theory, it needs to be renormalized at each order of the
expansion.  One introduces an effective operator in terms of the pion
field.  The most general Lagrangian density is then unique.  When expanded
in terms of $m_\pi / \Lambda$ and momenta/$\Lambda$, the free pion
Lagrangian is obtained to lowest order and pion-pion scattering is found
at the next order.  Although the agreement with experiment is quite good,
even at this order, it is improved by continuing the expansion through
the inclusion of higher order terms.\cite{Hol} 

    Recently the photoproduction of pions has received considerable
attention because it is a test of chiral perturbation theory.  At
threshold the production mechanism is dominated by the electric dipole
amplitude which is given by gauge invariance.\cite{E-W}  In chiral
perturbation theory the production amplitude is independent of the pion mass
and the pion-nucleon coupling constant.  At this order, the $\pi^0$
photoproduction from protons and neutrons vanishes.  But at the next order, 
an expansion in terms of $m_\pi^2/\Lambda^2$ and $(momenta/\Lambda)^2$
gives a finite value which agrees quite well with recent experiments.\cite
{Hol,Ber}.  Higher order calculations do even better.  We compare theory and
experiment in Table I. 

\begin{table} \caption{ Magnitudes of the amplitude for photoproduction
of pions in units of $10^{-3}/m_\pi^+$ to lowest (n = 1) and higher (n = 2
and 4) order in a chiral perturbation theory expansion compared to 
experiment.\protect\cite{Hol}}
\
\begin{tabular}{l|rrrr}
$Amplitude$          &n=1    &   n=2        &    n=4   &
     experiment     \\  \hline\hline
$\gamma p\rightarrow\pi^+n $  & 34.0  &26.4  &    &$28.4 \pm0.6$ \\
$\gamma n\rightarrow\pi^-p $  &-34.0  &-31.5 &    &$-31.8\pm1.2$ \\
$\gamma p\rightarrow\pi^0p $  & 0     &-3.58 &  -1.16  &$-1.31\pm 0.08$ 
\\
$\gamma n\rightarrow\pi^0n  $  &   0   & 0    & -0.44   & $\sim -0.4$
\end{tabular}
\end{table}

\subsection{Nucleon Structure}

	While we know that nucleons and mesons are composed of quarks and
gluons, the transition from a description of nuclei in terms of nucleons and
mesons to one in terms of quarks and gluons is still not understood.  Nor do
we fully understand the structure of the nucleons and mesons.  Progress has
been made by solving QCD numerically on a finite lattice rather than in
continuous space-time. 

	In an effective theory the gluon degrees of freedom do not appear
explicitly; in some models they are incorporated in the dressing of the
quarks, which then are called constituent quarks.  The constituent quarks (up
and down) have masses close to one third of the mass of the nucleon and thus
have small binding energies.  There are models of the nucleon made up of such
quarks, often treated non-relativistically and bound by harmonic or other
simple forces.  These models are amazingly successful in predicting the ratio
of the proton to neutron magnetic moments.  However, there have been a number
of authors who have pointed out that this result is not very model dependent.
Most effective theories (e.g., chiral perturbation theory) use almost massless
current quarks. 

	One of the original motivations for the study of nuclear structure was
to gain an understanding of the strong interaction.  Following this interest
in the structure of hadronic matter, nuclear physicists have become more and
more interested in a quantitative understanding of the structure of hadrons.
QCD provides the framework for such understanding.  For instance, QCD-based
constituent quark models not only can reproduce accurately the masses of
mesons with heavy quarks, but can also account for the main features of the
masses and electromagnetic decays of baryons with light quarks.  However, 
significant problems remain.  In particular, some signs and magnitudes of
strong decays of the higher mass nucleon resonances are poorly understood,
possibly because of the lack of a proper treatment of chiral symmetry.  The
lack of clear experimental evidence for particles that would correspond to
excitations of the gluonic field or ``hybrid" states of baryons that involve
involve gluonic excitations is an another outstanding puzzle. 

	In the last few years deep inelastic polarized electron
scattering on polarized H and D targets have provided insights into the
spin structure of the nucleon and have taught us that we do not yet fully
understand it.  For a given quark species (u,d, or s) the fraction of the
nucleon's spin that is carried by quark spins is defined as 

\begin{eqnarray} \label {1}
\Delta q = q\uparrow - q\downarrow + \bar{q}\uparrow - 
\bar{q}\downarrow,\\
\Delta \Sigma = \sum \Delta q, \\
\Delta u - \Delta d = g_A,
\end{eqnarray}
with arrows indicating spins parallel ($\uparrow$) and antiparallel
($\downarrow$) to the proton's spin and $g_A$ is the weak isovector axial
coupling constant.  The nonrelativistic ``naive" constituent quark model
predicts 

\begin{equation}  \Delta u = 4/3, \; \Delta d = -1/3, \;  \Delta s = 0 , \Delta \Sigma = 1, 
\end{equation} and 
\begin{equation}  g_A = \Delta u - \Delta d = 5/3. \end{equation} 

Eq. (5) turns out to be far from the truth.  Neutron beta decay yields 
$g_A=1.26$ and the deep inelastic scattering experiments show that only about
30\% of the spin of the proton comes from the quarks.  It is found 
that \cite {Ash} 

\begin{equation} \Delta u = 0.84 \pm 0.04, \qquad \Delta d = - 0.42 \pm 0.04,
\qquad \Delta s = -0.09 \pm 0.04, \qquad \Delta \Sigma = 0.33 \pm 0.08 .
\end{equation} 

	The experimental result came as a surprise and was called the ``proton
spin puzzle"; it implies that the major fraction of the proton spin comes from
the gluons and possibly from angular momentum.  However, $\Delta q$ is
generally evaluated in the infinite momentum frame and the nonrelativistic
quark model is for a nucleon at rest.  In terms of nuclear physics, the
angular momentum could come from pions coupled to quarks. 

	Another surprise was the relatively large ($\sim 10\%$ ) contribution
to the spin from strange sea quarks.  Again, in principle, this can be
understood in terms of low energy nuclear physics via the dissociation of
protons into strange baryons and mesons such as $\Lambda$ and $K^+$.
Antiquarks play a role because gluons can split into $q \bar q$ pairs and
because quarks may couple to mesons, particularly pions.  To the extent that
the Fock space contains pions, one can understand the excess of $\bar {d}$
over $\bar {u}$ in a proton since a one meson decomposition gives $ p =  p
\pi^0$ or $ n \pi^+$, with a $\pi^+ = u \bar{d}$.  Indeed, there is evidence
for an excess of $\bar {d}$ over $\bar {u} $ in the proton from inclusive
hadronic reactions with lepton pair production (Drell-Yan processes) and from
tests of the ``Gottfried sum rule", which follows from the assumption of
a flavorless sea of light quarks ($ u \bar{u} = d \bar{d}$).

\subsection{Nuclear Forces}

	The nucleon-nucleon force is basic to understanding nuclei and thus
has been of great interest for many decades.  Broadly speaking, potential
representations of the force are either purely phenomenological, or based on 
meson-exchange but with the parameters determined phenomenologically.  All
models have a one-pion exchange character at long range, which gives rise to a
spin-spin central potential and a tensor term.  The correctness of pion
exchange and its dominance at large distances is clear from the
nucleon-nucleon phase shifts at large angular momenta and from various
properties of the deuteron ground state, such as the non-zero quadrupole
moment.  Indeed, the strong tensor component of the pion-exchange force is a
unique feature that makes the solution of nuclear many-body problems
particularly challenging. 

	Some potential models represent the shorter range interaction by 
heavy-meson exchanges.  The Reid and the Urbana-Argonne potentials are 
examples of more phenomenological models, while the Nijmegen, Paris, and Bonn 
potentials are based more on meson exchange.\cite{Mach},\cite{E-W}  In the
last few years significant progress has been made in obtaining high-precision
fits to the elastic scattering data.  The ``Argonne $V_{18}$", the ``CD Bonn",
and several Nijmegen models all fit these data within the experimental
accuracy. 

	These modern potentials, coupled with recent advances in nuclear
many-body theory and in the capacity of computers now makes it possible to
understand the stability, structure, and reactions of light nuclei at a
microscopic level.  Three and four-nucleon systems are studied accurately in
both bound and scattering states by Faddeev and hyperspherical harmonic
methods.  For nuclei with up to 7 nucleons, quantum Monte Carlo methods have
been applied successfully.  Ground states, stable against breakup into
subclusters, are determined for $^{6}Li$ and $^{7}Li$, and their binding
and excited state spectra agree reasonably with the data.  In this work
three-nucleon forces are required; their strength is adjusted to reproduce the
binding energy of $^{3}H$ and to give a reasonable saturation density for
nuclear matter.\cite{Carl}

    	Chiral perturbative theories and other effective theories have also
been applied to the nucleon-nucleon problem.  The S-state scattering
lengths and effective ranges as well as the phase shifts up to momenta
of $\sim$ 300 MeV/c can be fit very well, as is illustrated in Fig. 1,
\cite {Kap} despite the difficulty of the anomalously large scattering lengths
in the $^1S_0$ state and the $^3S_1$ states.  Their effective potential
includes a contact interaction and pion exchange.  Related techniques with
effective chiral theories have been used by others.\cite{Jen}  It is
interesting that chiral perturbation theory allows one to show that 3-body
forces are smaller than two-body ones by the ratio $(p/\Lambda)^2$, where p
is a nucleon momentum and $\Lambda$, as before, is a QCD scale; for example,
if the two-body potential has an average strength of 20 MeV, then the
three-body one would have a strength of about 1 MeV.  The theory can also be
applied to charge independence- and charge symmetry-breaking forces and also
gives a strength ranking for these forces; \cite {VK1} e.g., one can show that
the charge symmetry-breaking force is weaker than the charge-independence
breaking one. 

	Very sensitive tests of the nucleon-nucleon interaction in bound
states are precise measurements of the radial distribution of nucleons in
nuclei for comparison with {\it ab initio} calculations.\cite{Car}  In many
cases electron scattering can measure these distributions directly.  However,
in the case of the deuteron which has unit spin, the orbital angular momentum 
0 and 2 contributions can be separated using polarized electron-deuteron 
scattering in the quantity $t_{20}$.  The current information on this 
quantity is illustrated in Fig. 2.  A measurement that should provide 
such information to momentum transfer $Q > 6 fm^{-1}$, or about 0.15 fm in the 
distances in the radial distribution of the deuteron, is among the early
experiments at CEBAF. 

	While these measurements demonstrate the wide validity of the hadronic
description of the deuteron, at the shortest distance scales the nucleon 
substructure does become important.  To date one nuclear reaction, the 
photodisintegration of the deuteron at large transverse momenta, into a
proton and a neutron, shows the behavior expected of coherently transferring 
the energy of the incident photon to the six constituent quarks in the 
deuteron.  Indeed, this reaction seems to show ``counting rule" behavior at 
considerably lower energies than was expected, suggesting a wider validity 
for quark descriptions.

\section{Nuclei}

	The successes in understanding the structure of nuclei are represented
by models such as the shell model and the collective model.  The
characterizations of the finite many-body system with complex forces between
the constituents is necessarily approximate. 

	The use of the Hartree approximation with simplified potentials
representing the nucleon-nucleon interaction, or even a field theory with
scalar and vector mesons (``quantum hadrodynamics") leads to a mean field in
which the nucleons move.  The theory can be extended to Hartree-Fock and to
include deviations from the mean field; it can be compared to experiment.  For
instance, it does well in reproducing the charge density determined by
electron scattering as is seen in Fig. 3. 

	Electrons are a great tool for precision studies of nuclei:  their
interaction is sufficiently weak that perturbation theory can be used, they
cause little distortion of the system, and their wavelengths can be made
sufficiently short to study both nuclei and nucleons in detail.  \cite {Die}
Electron scattering beautifully shows the single-particle structure of nuclei
in a mean-field description by measuring the properties of individual
shell-model orbitals and also by exploring the limitations of this description
in exploring the correlation effects that arise from the short-range part of
the nucleon-nucleon interaction.\cite{Pan}  This is clearly seen in the
proton knockout reactions illustrated in Fig. 4.  The consequences of 
these correlations are manifold.  They substantially renormalize the
single-particle mean field orbitals and appear to be the source of
high-momentum nucleons that are important in ``sub-threshold" production of 
mesons and other particles (e.g. anti-protons).  Such correlations can have an
effect on the mean-free paths of nucleons and other hadrons in nuclei and on 
mechanisms responsible for pion absorption in the nuclear medium.  But we are 
only at the beginning of being able to separate the effects of short-range 
correlations from other many-body effects. 

	Another perspective of the single-particle structure in nuclei with
many nucleons comes from an entirely different dimension.  While the successes
of the single-particle description in heavy nuclei are remarkable, the
description is tested almost entirely for the ``valence" orbitals and not for
the deeply bound states.  Such tests become possible by introducing a
different baryon into the nucleus that can settle into the lowest state
without violating the exclusion principle.  The most suitable baryon, because
of its relative stability, is the $\Lambda$, and the structure of the
deeply-bound states of the so-called ``hypernuclei" beautifully confirm the
single-particle description with the mean field modified to account for the
differences in the $\Lambda$-nucleon interaction.\cite{Chr} 

	Major advances have been made recently in exploring the structure of
nuclei under extreme, limiting conditions -- at very high angular momentum, in
approaching the limits of nuclear binding, and in temperature and energy
density (discussed in section V).  We discuss the first two in the
section below, as examples of recent developments in the field. 

        Major advances in recent decades have come primarily from novel and
substantially improved experimental techniques coupled with new theoretical
understanding.

\subsection {Nuclei at High Angular Momentum}

	In particular, accelerator developments have expanded enormously the
available beams.  With beams of heavy nuclei accelerated as projectiles it has
become possible to bring large amounts of angular momentum into nuclei.\cite
{Diamond}  For instance, with a 200 MeV $^{48}Ca$ beam incident on a target of
$^{120}Sn$ one forms a compound nucleus at high excitation energy, that first
rapidly decays (in $\sim 10^{-21}$ sec) by emitting particles (neutrons), and
then remains highly excited in a bound system, usually with high angular
momentum.  In a rotating reference frame, the excitation energy is not very
high -- it has to be measured with respect to the lowest energy that the
nucleus can have at this angular momentum (the so-called ``yrast line").  Under
the influence of centrifugal forces the lowest configurations in the nucleus
can be quite different from those at low angular momentum.  The shell structure
of nuclei becomes rearranged under the centrifugal effects of rotation and new
pockets of relative stability, in the potential energy surface of the nucleus,
may develop as a function of quadrupole deformation.  This then results in new
classes of nuclear states, with high deformation, in which many of the
nucleons have microscopic quantum numbers that are different from those of the
ground state.  

        The exploration of nuclear structure at high angular momentum has
shown that a few percent of the time fusion reactions populate states that
decay by electromagnetic cascades which show characteristic rotational bands
of remarkable simplicity as is shown in Fig. 5.  The energy spacings
in these bands correspond to nuclei with much higher deformations (2:1 axis
ratios) than those in the normal deformed bands that had been the basis of the
collective model of Bohr and Mottelson (typically ~1.3:1 axis ratios).  The
exploration of the properties of these {\it ``super deformed"} bands \cite 
{Janssens} has uncovered a great deal of structural information.  The precise
energies in these -- including the microscopic details of the small deviations
from the pattern expected of perfect rotors are reproduced with surprising
accuracy in several nearby nuclei, giving rise to the {\it identical band}
phenomenon \cite {Bak} as is illustrated in Fig. 6.  It seems that these
microscopic signatures carry over from one nucleus to another, without change,
in a way that has not been seen elsewhere in nuclear structure and is not
fully understood. 

        One of the most interesting features of superdeformed bands, mentioned
above, is the fact that even when these states are well above the yrast line,
they hardly mix at all into the higher density of states with more ``normal"
deformations.  There appear to be two distinct classes of states corresponding
to two minima in the potential energy surface, with different deformations. 
The superdeformed states are closely related to a class of states that
appeared in the 1960s in the study of delayed fission of very heavy nuclei.
The mixing between these states leading to fission and the ordinary states is
similarly inhibited.

        The discovery and study of these phenomena in nuclei at high-angular
momentum have become possible through major advances in the detection and
precision energy measurement of gamma rays with high-resolution germanium
diodes.  The size of the detectors, their anti-coincidence shields to suppress
the Compton-scattering background have now been developed to the point where
complete spheres of detectors are used to search for multiple coincident gamma
rays from a cascade.  This instrumental advance culminated in detectors such
as Gammasphere in the U.S. and Euroball in Europe.  The new experimental
information, in turn, has lead to major advances in the theoretical insights
into the structure of nuclei at the limits of large centrifugal stress. 

        Experimental work with these new instruments has also lead to other
new discoveries.  One of the most interesting of these are bands of states
connected by the emission of a sequence of electromagnetic quanta whose
energies increase in small smooth increments, very much as those for
rotational states.\cite{Mag}  However, unlike the electric quadrupole
radiations that are the earmark of transitions within rotational bands, the
gamma-ray transitions in these sequences are magnetic dipole in character. 
This came as a dramatic surprise.  The plausible explanation for these
magnetic transitions is that there are two, rather stable, configurations, one
for protons and one for neutrons, each of large angular momentum.  The
sequence of states with increasing angular momentum then correspond to states
where the two configurations simply are re-oriented to be more nearly parallel
and thus give higher total angular momentum, as the closing blades of shears. 
This explains the large magnetic dipole transitions, but the smooth dependence
to the sequence of energies is still very much a puzzle.  The phenomenon
suggests that some cooperative, collective features are present, though
theoretical understanding is still not complete. 

\subsection  {Limits of Binding}

        The 1950-1980 period saw the systematic exploration of the structure
of nuclei in and near the valley of nuclear stability through a variety of
techniques.  The limits of nuclear binding, where for a fixed number of
protons no more neutrons can be bound, (or for a fixed number of neutrons no
more protons), the so-called ``drip lines" were largely unknown, except for the
lightest nuclei.  The drip lines are of interest because nuclear properties
might change, especially near the neutron drip line.  They are also of
particular interest in various stellar processes where, in a hot environment,
a sequence of captures takes place rapidly.  With recent advances the
exploration of these limits has started in the 1990's. 

	One result along the drip lines is the observation of proton
radioactivity, where the nuclei are literally dripping protons because their
binding is insufficient but the Coulomb barrier retards their emission.  At
present, these have been identified in a number of elemental isotopic
sequences, from Co to Bi, and the structure of these nuclei at the proton drip
line is beginning to be explored.\cite{Woo} 

	Another result is in the limit of neutron binding where much less is
known, because this regime is much more difficult to reach in laboratory
experiments.  Since neutrons are are not subject to the Coulomb force, loosely
bound neutrons result in long tails in the density distributions that fall off
exponentially -- but with the exponent decreasing with binding.  Thus the
neutrons will reach far beyond the proton distributions in very neutron-rich
nuclei.  Such a separation between neutrons and protons, might in turn cause
some qualitative changes in nuclear properties.  For instance, it has been
suggested that the spin-orbit term may be substantially reduced in such
nuclei, and this would cause a change in shell structure that would be very
interesting to observe experimentally.  Such a change could also have serious
consequences on the rapid neutron capture, r-process, in explosive stellar
nucleosynthesis. 

	The best current example of a nucleus with diffuse neutron excess is
in a very light nucleus $^{11}Li$ with three protons and eight neutrons.  Here
experiments show clearly how the last two, very loosely bound, neutrons form a
diffuse tail, a ``halo" around the protons.  Thus the interaction radius of
this nucleus is substantially larger than that of other Li isotopes as is 
shown in Fig. 7 and the structure of, what would be the electric
dipole giant resonance in other nuclei, is substantially different here, as is
the momentum distribution.\cite{han}  The further exploration of very
neutron-rich nuclei, beyond the very light ones, requires major new advances
in experimental techniques and facilities. 

	Another limit is being explored in the size a nucleus can reach in its
total mass, or nucleon number.  Here ingenious improvements in experimental
techniques permit the production of heavier and heavier new isotopes and
elements -- beginning to approach the region where calculations predict that,
because of the stabilizing effects of shell structure, a new island of
relatively stable {\it ``superheavy"} nuclei should occur.  A few nuclei of the
new element with Z = 112 have been produced -- near one such possible island
with Z = 114, but further from another suggestion with Z = 126.  The results may
in fact indicate a relatively stable bridge leading to a more stable island.
Very heavy atoms are also of interest because of their atomic structure, to
explore the rapidly increasing effects from the large Coulomb field in
approaching the limits of vacuum polarization as well as the increasing
relativistic effects.\cite{superheavies}

\subsection {Hadrons in the Nuclear Medium}

	A crucial assumption in most of our many-body descriptions of nuclei
is that nucleons and other hadrons do not change in the nuclear medium.  Deep
inelastic scattering of electrons on nuclei gave evidence that nucleons and
their quark structures are altered somewhat when they are placed in nuclei.
\cite{Gee}  This is called the EMC effect after the group that first
discovered it.  Their finding has caused interest in studies of these changes
both experimentally and theoretically and has raised important issues, more
generally, about how the properties of hadrons change in the hadronic medium
of a nucleus.  The changes can be investigated at the quark level in high
momentum-transfer reactions and at the hadron level in both electron
scattering and heavy-ion collisions.  Using electrons, the elastic form
factors of the proton and neutron inside nuclei have been compared to those of
free protons and neutrons in deuterium.  Intriguing differences in the ratio
of the magnetic to electric properties of the proton in light nuclei have been
observed, and this is an active area of investigations. 

	On the other hand, there is clear evidence for changes in the 
effective nucleon-nucleon force in the nuclear medium.  The challenge is to 
distinguish ``normal" many-body effects from changes in the hadronic 
substructure.  For instance, how do virtual pions from pion exchange manifest 
themselves in a change of the sea antiquark distribution of nucleons in 
nuclei?  This is being examined at the hadronic level in looking for pions
knocked out by electrons, in Drell-Yan processes that are directly sensitive
to antiquarks, and in looking for pionic modes excited in proton scattering. 
If the structure and properties of the mesons change in the nuclear medium,
there may be profound implications for the effective internucleon forces. 

	These changes involve not only pions but also vector mesons.  Theorists
have suggested that the masses of the latter should decrease\cite{GB}\cite{Ko},
but there remains controversy about this suggestion.  The width of the
$\rho$ resonance is also likely to be affected.  These alterations can be
sought by searching for the leptonic decays of the vector mesons produced in
heavy-ion collisions and studying their leptonic decays within the nucleus. 
Such experiments are being undertaken.  Many of the changes of hadrons are not
large, but the methodology seems to be at hand to seek them out and the
consequences could have a profound impact throughout nuclear physics. 

	When a colorless hadron is produced in a high momentum transfer
reaction, the hadron is small at birth.  Due to color screening of the quark
components of this small hadron, its mean free path in the nucleus is large
and its final state interactions small.  This phenomenon is called color
transparency,\cite{Mi} and should be visible in quasielastic (e,ep) and 
(p,pp) reactions on a nuclear target.  At present the evidence for nuclear
transparency is ambivalent and further experiments are required to tie the
effect down.

\section{Nuclear Astrophysics}

        Nuclear physics plays a key role in the processes that take place in
the universe, from the Big Bang on to the energy production in stars.  The
synthesis of the chemical elements in the universe is the result of the
various nuclear reactions that take place in different stellar environments.
The Big Bang produced mostly protons and some of the lightest elements.  When
stars are formed from these remnants of the Big Bang, their matter is
heated as the star contracts under gravity.  In the hot star, nuclei run
through cycles of nuclear reactions and hydrogen is gradually converted into
helium, as in our sun.  Somewhat hotter stars will form carbon, and as the
carbon cycle described by Bethe becomes important more hydrogen is converted
into helium.  Further heating will cause captures of protons and alpha
particles beyond the carbon cycle.  Under appropriate conditions these
reactions will produce elements up to about mass 56.  To get further in mass,
neutron capture is essential -- and this can happen slowly in the ``s-process"
or explosively in the cataclysmic ``r-process".  All the elements heavier than
iron in our world were produced in such stellar environments.  Below we mention
only a few of the key developments in our understanding of the recent past --
following the major insights in this field by Bethe, Fowler, and others in
identifying how nuclear processes determine the evolution of matter in the
universe, in describing energy production in the Sun and in explaining the
formation of the elements from the Big Bang through the stages of stellar
evolution. 

\subsection{Solar Neutrinos}

	One of the intriguing developments of the past decades has been the
study of neutrinos from the Sun.  Other than the heat radiated, the
neutrinos are the one accessible observable product from the chain of
nuclear processes that take place in the interior.  It was noted early that
the number of neutrinos detected on earth was too small, given that the energy
output of the sun is known accurately and thus the number of nuclear reactions
leading to neutrinos is also known.  The pioneering experiments of Davis and 
colleagues with a chlorine detector in the Homestake mine in South Dakota were
sensitive primarily to the highest energy neutrinos, those from the decay of
${^8}B$.  The observed neutrino flux was about a third of that expected, and
in spite of extensive measurements and remeasurements of the nuclear
parameters relevant to the solar processes and to the detection scheme, the
discrepancy remains today as is shown in Fig. 8.  

	More recent experiments with gallium as the detecting material, in
Europe and Russia are sensitive to much lower energy neutrinos.  They also
find substantially fewer neutrinos than expected, as do experiments in which
the high-energy neutrinos are detected more directly in large water detectors
in Japan.  The combined impact of these very different experiments has already
been profound.  The deficiency in the number of electron neutrinos might have
its origins in the possibility that neutrinos have finite mass and that they
oscillate between the originally emitted electron neutrinos and neutrinos of
other flavors, with this oscillation enhanced by their passage through the
dense matter of the Sun.  Whether this is indeed the case will be tested in
experiments to be carried out in the coming years.\cite{solar}

\subsection {Supernovae}
	Among the most spectacular events in the universe are the supernova
explosions that occur under the right circumstances during the evolution of 
sufficiently massive stars.  The dynamics of such an explosion involves a
complex interplay between nuclear properties and gravity the weak interaction.
 Supernovae play by far the dominant role in the synthesis of heavy elements. 
The decay in the light curve of a supernova, shown in Fig. 9, is
governed predominantly by the decay of $^{56}Ni$ -- the progenitor of
$^{56}Fe$, the most tightly bound nucleus and the most abundant element
constituting the earth.  The enormous flux of neutrinos from supernovae was
evidenced by the dramatic detection of a neutrino pulse from the supernova
SN1987A.  Recent theoretical work has shown that the shock front of neutrinos
interacting with nuclear matter plays a major role in the dynamics of
supernova explosions. 

\subsection{Neutron Stars}

	A typical neutron star has about 1.4 solar masses compressed in a
sphere of ~10 km radius.  More than 500 neutron stars have been detected in
our galaxy, most as radio pulsars and some at optical and x-ray wavelengths. 
A dozen neutron stars are found in close binary pairs, which has allowed their
individual masses to be measured very accurately through orbital analysis. 
Binary neutron star coalescence events may be the source of observed
extra-galactic gamma-ray bursts, and our best hope for directly detecting
gravitational radiation. 

	The structure of a neutron star is a consequence of the interplay
between all interactions:  the strong nuclear force, electroweak interactions,
and gravity, with significant corrections from general relativity.  The surface
is metallic iron, which is the most stable form of matter at zero temperature
and pressure.  Underneath is a lattice of nuclei that grow progressively
larger, more neutron-rich, and more tightly packed as the density increases
with depth, since it is energetically favorable to capture electrons on
protons to reduce the kinetic energy of the electron gas.  At about $4\times 
10^{11} g/cm^{3}$ matter density, neutrons start to leak out of the nuclei,
forming a low-density neutron superfluid in the intervening space.  As the
density increases further, around $10^{14} g/cm^{3}$, various unusual shapes
of nuclei may occur, with transitions from spheres to rods to sheets to tubes
to bubbles.  Eventually at a density around $2.7\times 10^{14} g/cm^{3}$, or
normal nuclear matter density, the nuclei dissolve into a uniform fluid which
is over 90\% neutrons, 5-10\% protons and an equal number of electrons to
preserve electrical neutrality, all in beta-equilibrium.  At successively
higher densities muons, and perhaps pions and/or kaons may be present.  At
high enough densities, quark matter will probably appear, perhaps initially as
bubbles in the nucleon fluid. 

	Progress in characterizing the nucleon-nucleon interaction has been
key to our increased understanding of dense nucleon matter and consequently of
neutron star structure.  If neutrons were noninteracting, the maximum neutron
star mass stable against gravitational collapse to a black hole would be 0.7 
solar masses.  However, many neutron stars have been observed with masses
1.3-1.6 times the solar mass, so the role of nuclear forces is clearly
important.  Recent work has set the minimum upper limit on neutron star mass
at 2.9 solar masses, helping to further refine the observational boundary
between neutron stars and black holes.  Observations of quasi-periodic
oscillations in binary x-ray sources may also soon lead to limits on neutron
star radii, which will give an even tighter constraint on the dense matter
equation of state.

\subsection {Nucleosynthesis and Reactions with Unstable Nuclei} 

	The nuclei of the chemical elements are formed in the very hot
environment inside stars.  At higher stellar temperatures, these processes
occur sufficiently fast that the nuclei involved are themselves short-lived.
New tools are being developed in nuclear physics that will help meet the
challenge to determine properties of these short-lived nuclei, including the
cross sections that are likely to be most important in astrophysical contexts.
 In addition, there are some expectations that the general outlines of nuclear
structure may change near the limits of binding, where the path of explosive
nucleosynthesis takes place. 

	The enormous improvements in the observational techniques of astronomy
and astrophysics inevitably will require better quantitative understanding of
the nuclear processes that yield energy in the universe and form an intriguing
interface with nuclear physics.

\section{Matter at High Energy Densities}

        One of the areas of intense interest is associated with the physics at
very high densities, where the description of matter in terms of quarks and
gluons contained in individual hadrons must break down.  Calculations based on
QCD suggest that when high densities and temperatures are reached in a volume
large compared to that of a typical hadron, a transition to a state of matter
will occur where the quarks are no longer confined to their individual hadrons
but can move freely within the larger volume. 

	A schematic phase diagram is shown in Fig. 10 illustrating
this transition and its relationship to the evolution of the universe
following the Big Bang.  To explore this state, experiments have been carried
out, first at the Bevalac at Berkeley, then at increasingly higher energies at
the AGS in Brookhaven and the SPS at CERN, and soon at a new collider, the
RHIC (Relativistic Heavy Ion collider) facility at Brookhaven that is to come
into operation shortly.  Since the confinement of quarks in hadrons is one of
the key features of the strong-interaction world, this deconfinement, to a
volume larger than that of a hadron, is of course of intense interest to
physicists.  A key question is how this deconfinement might be observed in an
unambiguous fashion. 

        The collisions between two heavy nuclei at ultrarelativistic energies
produce many thousands of fragments.  Their detection, identification, and
characterization is a formidable challenge to experimentalists.  The first
question is whether the large densities in energy and baryon number are indeed
obtained in such collisions -- whether their kinetic energy is absorbed or,
whether the two nuclei primarily just pass through each other.  This may be
deduced from the measurements of the transverse momentum carried by the
products of the interaction.  Such "stopping" studies at the AGS, at
laboratory energies up to 14 GeV per nucleon show clearly that the kinetic
energy is absorbed, and that high temperature and high baryon density are
indeed created.  This remains true in work at CERN up to laboratory energies
of 200 GeV per nucleon.  As yet, there is no unequivocal evidence in these
data for a phase transition.  The energies at RHIC will be an order of
magnitude higher -- and it is expected that this will lead to the formation of
a high-temperature low baryon-density system well past the expected transition
point. 

        How the properties of this transitory state of matter can be deduced
from the data is the subject of intense discussions between experimenters and
theorists.  Some of the possible ``signatures" discussed and explored in
current experiments are the suppression of the production of $J/{\psi}$ mesons
due to screening of the charm-anticharm interaction by the surrounding quarks
and gluons, an increase in energetic leptons in radiation from the early
stages of the system, the modification of the width and decay modes of
specific mesons, such as the ${\rho}$ and ${\phi}$, enhancement in the
production of strangeness, the attenuation of high energy jets, etc.  A change
in the state of matter should first show up in a clear correlation between a
number of such signatures. 

        Another possibility at high energy density is that the intrinsic
chiral symmetry of QCD may be restored.  This would show up through
modifications in the masses of mesons.  Possibly, non-statistical fluctuations
in the distributions of pions, might signal the production of a so-called
``disoriented chiral condensate".  The major point is not the specific
scenarios in such a complex environment, but that these energy densities will
bring an unprecedented new regime of matter under experimental scrutiny where
our current pictures will necessarily have to undergo radical changes.\cite{Har}.

\section{Tests of the Standard Model}

	Most of the tests of the standard model of the strong and electroweak
interactions at nuclear energies have involved semi-leptonic interactions,
particularly electrons and nucleons, and rare decay modes, primarily of kaons.
Here we describe some of the semi-leptonic tests. 

\subsection {Beta Decay}

The study of superallowed Fermi beta decays ( parent nucleus of spin/parity
$J^{\pi}=0^{+} \rightarrow$ daughter $0^{+}$) in the same family of
isospin (isospin multiplet)  permit an important test of the standard model to
be carried out via the unitarity of the matrix that describes the weak
interaction connecting the various quarks:  the Cabibbo-Kobayashi-Maskawa (CKM)
matrix \cite{Tow}.  The matrix element $V_{ud}$ connecting up and down quarks
is by far the largest one in the unitarity of $$ U\equiv \mid V_{ud} \mid ^{2}
+ \mid V_{us} \mid ^{2} + \mid V_{ub} \mid ^{2}  = 1 $$ 

\noindent Here $V_{us}$ and $V_{ub}$ connect the up quark with the strange and
bottom quarks, respectively.  The precise measurements of superallowed
transitions together with radiative corrections and removal of charge
dependent nuclear effects allow one to determine $V_{ud}$ to better than
$10^{-3}$.  In addition, these measurements, shown in Fig. 11 demonstrate
that CVC holds to $\sim 4 \times 10^{-4}$. \noindent A straightforward
analysis of the experiments, including a recent $^{10}C$ experiment, gives
$V_{ud} = .9740 \pm. 0006$.  Together with the measurements of $V_{us}$ and
$V_{ub}$, one then obtains $U = .9972 \pm .0019$.  There remain uncertainties,
particularly in the charge-dependent nuclear effects; it has been proposed
that these corrections can be approximated by a smooth Z-dependence.  In that
case $U = .9980 \pm .0019$.  However, better calculations of this
charge-dependence remains to be carried out. 

\subsection{Double Beta Decay}

	The decay $(A,Z) \rightarrow (A,Z+2)+ e^{-}+ e^{-} +\bar{\nu}
+\bar{\nu}$ is expected in the standard model and has been seen in several
nuclei ($^{82}Se$,\, $^{100}Mo$,\, and $^{150}Nd$) with half lives of about
$10^{20} y$, consistent with the standard model.\cite{MM}  Searches for the
no-neutrino decay mode, important for determining whether $\nu$'s are massive
and of the Majorana (neutrino and antineutrino are identical) type, have
been continuously improved.  The present lower limit on the half life is
$3\times 10^{24} y$ for $^{74}Ge$.  Improved experiments hope to find this 
mode of double beta decay. 

\subsection{Semi-Leptonic Parity-Nonconservation Studies} 

	The finding of strangeness in the nucleon led to considerable
experimental and theoretical activity to understand it.  Recently, in an
ongoing experiment (SAMPLE) at MIT, the weak interaction of the electron and
proton is being used to investigate strangeness in the nucleon.  As in all
parity violation experiments, it is the interference of the weak interaction
with the electromagnetic one that is being detected by searching for a
parity-odd signal such as $<\vec{j}> \cdot \vec{p} $, where $\vec{p}$ is the
incident momentum of the electron and $<\vec{j}>$ is its polarization.  The
presence of strangeness in the nucleon can modify the momentum dependence of
the form factors, but it can also add two new unconstrained ones, a vector
magnetic form factor and an axial (isoscalar) form factor.  The results
to date \cite {SAMPLE} are inconclusive, but do not indicate any strangeness 
(within large errors), as shown in Fig. 12.  Further experiments
are planned.   

	Other precision parity-violating (pv) studies of the weak interactions
of electrons and nuclei have been carried out with atoms.  Despite their being
at lower momenta, where the pv effects are smaller, $\sim 10^{-11}$, these
experiments have reached the incredible precision of $1/2\%$ in the
parity-violating asymmetry.  At the present time theoretical errors are at the
level of $\sim 1\%$.  At this level of precision the atomic experiments provide
meaningful tests of the standard model.  The dominant weak interaction term is
$a_\mu V^\mu$ where the lower case $a_\mu$ is the axial current of the
electron and $V^\mu$ is the vector current, which is coherent over the
nucleus.  The effective charge, the weak equivalent of the electrical charge in
this case, is \begin{equation} \label Q Q_W = (1-4 \sin^2\theta_W)Z-N \; ,
\end{equation} where $\theta_W$ is the Weinberg angle, $\sin^2\theta_W \simeq
0.23$.  $Q_W$ is large for heavy atoms.  The measurement on Cs, a one-valence-
electron atom, at the $0.5 \%$ level, gives $Q_W = -72.35 \pm 0.27_{exp} \pm
0.54_{th}$.\cite{CSW}  This limits deviations from the standard model. 

	The term $v_\mu A^\mu$, where $v_\mu$ is the weak vector current of
the electron and $A^\mu$ is the nuclear axial current, is much smaller than
$a_\mu V^\mu$ because for the electron $v^\mu \propto (1-4\sin^2\theta_W)
\sim0.1$ and only a single nucleon contributes to $A^\mu \propto
<\vec{\sigma}>$, the nuclear spin.  Thus, the asymmetry is reduced by $\geq
500$.  The atomic measurements of this term make use of the hyperfine
structure, which is due to the nuclear spin.  This term has not yet been
detected because it is hidden by the stronger nuclear anapole moment, a
toroidal axial current of the nucleus coupling to photons.  That is, this is a
parity-violating coupling of the photon to the nucleus.  The recent
measurement in atomic Cs at the 1/2$\%$ level\cite {CSW} has discovered the
nuclear anapole moment, the first anapole moment observed for such a
microscopic system. 

	The experiment is of particular interest because it is sensitive to
the weak inter-nucleon force caused by neutral currents.  This force has not
been found in pure nuclear experiments and the upper limit found there (in
$^{18}F$) is at least a factor of three below that deduced from the anapole
measurement, and also from a theoretical quark model.  It remains to be seen
whether this is an experimental or theoretical problem. 

\subsection{The Non Leptonic Weak Interaction}  

	The nonleptonic weak interaction is of interest because it is the
least well understood one due to the strong interactions of all particles
involved.  The initial knowledge came from the weak decays of strange mesons
and baryons.  Even here, there remains the problem of fully understanding the
ratio of parity-non-conserving to parity-conserving amplitudes in the decays
of hyperons.  Due to the change in flavor, only charged currents contribute to
these decays in the electroweak theory.  With the advent of precision nuclear
experiments it became possible to study the weak interactions of nucleons by
means of parity-violating asymmetries with polarized beams.  These experiments
do not probe the standard model as much as as our understanding of the
structure and weak forces of the nucleons.  The asymmetry comes about from the
interference of the weak and strong forces.  Experiments in $pp$ scattering and
in light nuclei have provided the most reliable information.\cite {EGA}  The
weak neutral currents have yet to be seen in non-leptonic weak interactions.
They are the primary source of the isospin-changing $\Delta I$ = 1 interaction,
which arises from $\pi$ exchange with one weak ($f_\pi$) and one strong pion-
coupling to nucleons.  Since there is only an upper limit on the asymmetry in
$^{18}F$, which is determined by this mechanism, neutral current effects have
not yet appeared.  Further measurements are planned for $f_\pi$, e.g., in low
energy polarized neutron capture by hydrogen, e.g., $n + p \rightarrow d +
\gamma$. 

\subsection {Time Reversal Invariance}
 	Despite the finding of CP violation in 1964, over 30 years ago, we
still do not have any definitive theory of time reversal noninvariance.  This
is not due to a lack of effort.  The only system where CP violation (and by
implication T violation) has been found is in the $K^0$ and $\bar{K}^0$
system.  To date, the most sensitive searches for time-reversal-invariance
breaking are those for an atomic electric dipole moment of a neutron \cite
{Smi} or atomic $^{199}Hg$ \cite {For}; these tests are sensitive to
simultaneous violations of parity and time reversal invariance.  No finite
time-reversal violation effect has yet been seen, but the upper limits keep
decreasing and have already ruled out some models of CP violation. 

\section {Facilities and Sociology}
	The development of nuclear physics, particularly in the latter part of 
the century, has required major investments in facilities for acceleration of
the required probing beams and the detection systems required for experiments.
While the scale of these has not been the same as those of particle physics,
they have been sufficient to require planning and priority choices, carried 
out primarily through the Nuclear Science Advisory Committee, with broad
participation by the scientific community. 

	Out of these recommendations new facilities have emerged:  CEBAF and
RHIC are the two major examples, but a number of smaller facilities have been
constructed as part of this planning process; also substantial U.S. 
participations in international collaborations have occurred.  At the same
time, this planning process has also phased out major facilities in order to
make new initiatives feasible under a constrained budget.  

\section {Outlook}

	The study of the structure of nuclei, of hadronic matter, started less 
than 100 years ago.  Enormous advances have been made in this time.  But much 
work remains.  Even the simplest building blocks of hadronic matter that we
have in our world, the proton, neutron, or pion are structures that are
incompletely understood.  So are the interactions between them, the
quantitative features of the forces that hold nuclei together.  The properties
of nuclei as observed experimentally are understood in the framework of
approximate models, but the more fundamental reasons for the validity of many
of these models are not well understood -- nor can we reliably extrapolate
these properties to limiting conditions, whether in the limits of 
stability, or the limiting energy densities of matter.

	The concerted efforts of nuclear physicists, theorists and
experimentalists is needed to pursue these areas of knowledge into the 21st
century. 

\section {Acknowledgments}

	The authors would like to thank their colleagues G. Bertsch, D.
Geesaman, W. Haxton, R. Janssens, G. Miller, and R. Wiringa for helpful
discussions and advice in connection with the preparation of this article. 

        This research was supported by the U.S. Department of Energy, Nuclear 
Physics Division, under Contract W-31-109-ENG-38 and Grant DE-FG06-90ER40561.

\newpage

FIGURE CAPTIONS

Fig. 1
	The phase shift $\delta$ for the $^1S_0$ channel.  The dot-dash curve
is a one parameter fit in chiral perturbation theory at lowest order.  The
dotted and dashed curves are fits at the next order in the expansion; the
dashed one corresponds to fitting the phase shift between $0 \leq p \leq 200
$MeV, whereas the dotted one is fit to the scattering length and effective
range.  The solid line corresponds to the phase shift obtained from a partial
wave analysis carried out by the Nijmegen group.\cite{Kap} 

Fig. 2 
	The results of polarization measurements from electron scattering from
the deuteron.  The quantity $t_{20}$ from a variety of experiments is shown by
different symbols, with predictions of different theoretical models of the
nucleon-nucleon interactions drawn by lines.  New measurements from CEBAF 
should help to better distinguish between the models. 

Fig. 3
	The points show the difference in charge densities between $^{206}Pb$
and $^{205}Tl$ and show the oscillations due to the radial nodes in the
wavefunction of the last proton \cite {Pan}.  The line is a Hartree-Fock
calculation of the same difference. 

Fig. 4
	Demonstration of high-momentum components in the nuclear wave
function.  Transition densities from electron scattering knocking out a proton
from the single-particle states in the doubly-magic nucleus $^{208}Pb$.  The
inset shows the spectrum of hole states.  The solid line represents the
transition density calculated using a mean field for the protons, the
dot-dashed curve includes the effects of short-range, high-momentum
correlations. \cite {Pan} 

Fig. 5
	A ``super-deformed" rotational band in $^{152}Dy$ showing the gamma-ray
transitions between members of the band.  The constant increments in gamma-ray
energies are characteristic of a band that follows the symmetry of a very good
quantum-mechanical rotor.  The variation in intensity of these gamma rays
reflects the angular momentum distribution in the population of the band in
the fusion-evaporation reaction that was used.  The peaks that are not
indicated by arrows correspond to states of lower energy of ``normal"
deformation, populated after the decay out of the super-deformed band.\cite{twi} 

Fig. 6
	Similarities between super-deformed bands in nuclei in the same
vicinity.  The quantity plotted is the mean percentage difference in
$E_\gamma(J+2\rightarrow J)-E_\gamma(J\rightarrow J-2)$ between different
bands.  About sixty bands within the mass 150 region and fifty bands within
the mass 190 region are compared.  Note that in both regions there is a large
excess of pairs of bands for which this quantity differs from zero by less
than 2\% -- these are the ``identical bands" -- others differ by up to 20\%. 
This identical reproduction of bands in different nuclei is not yet fully
understood. 

Fig. 7
	Interaction radii of Li (3 protons) and Be (4 protons) nuclei with
carbon derived directly from total cross section measurements.  The line
represents a smooth $A^{1/3}$ dependence of the radius.\cite{Tan} 

Fig. 8
	The solar neutrino signal expected with three different detection
methods that are sensitive to different neutrino energies.  The light bars
indicate the expected neutrino yield, in appropriate units, from the various
sources in the solar cycle.  The solid bars represent the observed signals
from the corresponding experiments.  The shaded areas represent the
uncertainties. 

Fig. 9
	The logarithm of the light intensity from the supernova SN1987A as a
function of time after the supernova explosion.  Most of the intensity for the
first two years comes from $^{56}Ni$ and its daughter $^{56}Co$.  The
calculated contributions from other radioactive nuclei is also shown. 

Fig. 10
	A qualitative phase diagram for hadronic matter showing the transition
to a deconfined quark-gluon state at high temperature and high density.  The
paths followed in the early universe and in the interior of neutron stars is
also indicated.  $\rho_{0}$ is the density of normal nuclei.  Collisions in the
relativistic heavy-ion collider RHIC will explore this phase diagram in
detail.

Fig.  11
	Beta-decay transition probabilities (Ft values) for nine superallowed
Fermi $\beta$-decays and the best least-squares one-parameter fit, plotted as
a function of the proton number of the final nucleus.\cite{Tow} 

Fig. 12
	Results for the parity-violating asymmetry measured in the SAMPLE
experiment in the 1995 and 1996 running periods.  The hatched region is the
asymmetry band (due to axial radiative corrections) for $ F_2^s =0$. 

\end{document}